# Co-Emulation of Scan-Chain Based Designs Utilizing SCE-MI Infrastructure

Bill Jason Tomas[1], Yingtao Jiang[2] and Mei Yang[2]

[1]Cadence Design System, Inc., San Jose, CA, USA
[2]Department of Electrical and Computer Engineering, University of Nevada, Las Vegas
Las Vegas, NV, USA

**ABSTRACT**

*As the complexity of the scan algorithm is dependent on the number of design registers, large SoC scan designs can no longer be verified in RTL simulation unless partitioned into smaller sub-blocks. This paper proposes a methodology to decrease scan-chain verification time utilizing SCE-MI, a widely used communication protocol for emulation, and an FPGA-based emulation platform. A high-level (SystemC) testbench and FPGA synthesizable hardware transactor models are developed for the scan-chain ISCAS89 S400 benchmark circuit for high-speed communication between the host CPU workstation and the FPGA emulator. The emulation results are compared to other verification methodologies (RTL Simulation, Simulation Acceleration, and Transaction-based emulation), and found to be 82% faster than regular RTL simulation. In addition, the emulation runs in the MHz speed range, allowing the incorporation of software applications, drivers, and operating systems, as opposed to the Hz range in RTL simulation or sub-megahertz range as accomplished in transaction-based emulation. In addition, the integration of scan testing and acceleration/emulation platforms allows more complex DFT methods to be developed and tested on a large scale system, decreasing the time to market for products.*

**KEYWORDS**

*Design verification, emulation, SoC, scan-chain, SCE-MI, FPGA.*

## 1. INTRODUCTION

To gain controllability and observability into a digital system, designers utilize the 'Scan-Chain' testing methodology [1]. There are, however, a number of issues during verification of scan chain systems using RTL simulators. RTL simulation test time is dictated by the complexity of the test bench and the design, and the workstation CPU. Another factor is the RTL simulation tool used by the verification team. There are two types of RTL simulators used for verification: event-based and cycle-based. In this paper, we will focus our discussion on the use of event-based simulators. An event-based simulator updates whenever an event occurs, be it combinatorial or sequential [13]. Event-based simulators can capture transitions through a combinatorial datapath, which may not be aligned with a clock edge allowing users to capture issues such as glitches. Events are placed in a timing queue, which is evaluated in the order the events are placed [14]. In terms of CPU usage, a majority of the workload is utilized to update the events queue. Since large SoC designs can contain thousands of registers, simulation time can take hours or even days depending on the system complexity. Also, a bulk of the transitions occurs between the combinatorial paths between all scan registers. To speed up this process, one may need to utilize a development platform so that the scan chain can be placed into hardware and removed from the simulation environment.

There are a number of issues and use cases when utilizing different development platforms for scan testing. Each platform has advantages and disadvantages in terms of speed, cost, debug capabilities, and bring-up. Fig. 1 showcases the various platforms with the different HW/SW levels which is applicable for verification [15].

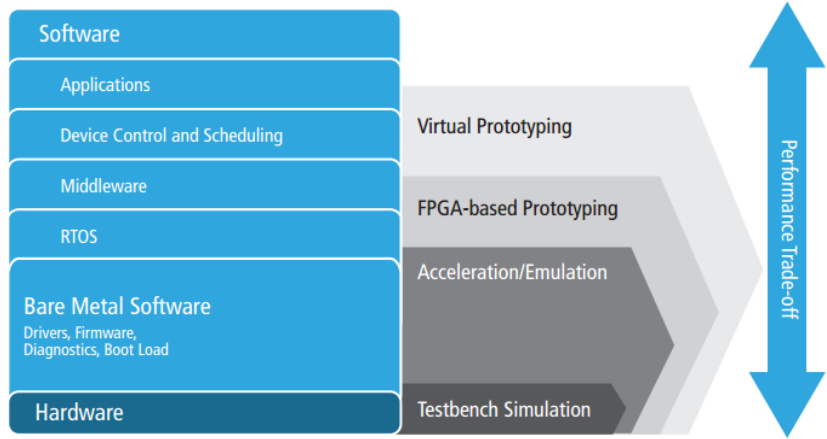

Fig. 1 Verification Techniques for Different Levels of Logic

Depending on which platform is used for hardware and software development, the debugging capabilities for verifying scan chain implementation differs. The best option would be to utilize a transaction-based emulation platform, since the scan chain will implemented on hardware, and can be verified utilizing transactions from high-level testbenches. Another option available by some FPGA vendors is an FPGA-based emulation platform, which combines the capabilities of an FPGA-based prototyping board with a transaction-based emulator. With this test environment, virtual models available from virtual prototyping solutions can be directly connected to an emulator workstation via software TLM libraries developed by emulation vendors. With high-level models and verification constructs, the scan chain can be verified utilizing timing accurate behavior made available by hardware.

In what follows, Section 2 discusses the advantages and disadvantages of different platforms that can be used for scan chain verification. Section 3 briefly introduces the SCE-MI interface and protocols. Verification of scan chain using modified SCE-MI interfaces will be detailed in Section 4, and the results are reported in Section 5. Finally, Section 6 concludes the paper.

## 2. HW/SW DEVELOPMENT PLATFORMS AND DEBUG

Generally, there are four development platforms for scan testing. We will discuss use cases and how scan chain verification can be accomplished use these platforms.

### 2.1. Testbench (RTL) Simulation

This is the main way of verifying design blocks utilizing Verilog, VHDL, or SystemVerilog. For scan testing, a user can simply initialize test vectors established from fault simulation, and use HDL constructs to feed the data serially into the scan data input. In a similar fashion, data from the scanout can be outputted to a text file to be analyzed at a later time or compared to against a golden set.

The biggest advantage of using RTL simulation is that all signal transitions are presented to the user through waveform viewers, schematic editors, and various debug tools. If there is an incorrect value in the scanout output, user can observe all signals in the chain as well as the current state of all scan registers. This allows the user to quickly debug a broken scan chain. Although this is the lowest cost of all platforms, large SoC designs incur very long simulation cycles. In this manner, a scan chain cannot quickly be verified on the system level because bugs can only be found at the end of the simulation run.

## 2.2 FPGA-based Prototyping

Prototyping platforms, such as the Aldec HES-7, utilize FPGA place and route process (P&R) with a generated bitstream which is implemented onto FPGA memory [16]. RTL is synthesized and mapped to FPGA primitives, which then runs through a placer which connects the primitives with FPGA interconnect.

Most prototyping boards include peripherals and connectors (e.g., PCIe, RS232, and USB) which the user can map design IO to a target device. Since the entire design is running in hardware, it is capable of running at faster clock frequency than RTL simulation. The disadvantage of using an FPGA-based prototype is that there is little debug capability since signals cannot be observed using a waveform viewer [16]. On-board logic analyzers, such as Xilinx ChipScope Pro and Altera SignalTap, mitigate the issue, but with a limited number of bits to be sampled, and they do not provide a complete debug environment. Debugging a scan chain is difficult when implemented onto an FPGA, since users can only observe the scanin and scanout data. If a scan register contains an incorrect value, it must propagate through the entire chain before being observed at the output.

## 2.3 Acceleration/Emulation

Acceleration and emulation platforms combine the debugging environment from RTL simulation with the speed of an FPGA prototype. In these systems, a workstation with an RTL simulator directly connects to a FPGA prototype via high-speed interfaces (e.g., Ethernet and PCIe). Emulation teams develop hardware drivers for the physical interface and a software library to establish communication between the hardware and software. From this point, users can utilize two different modes of operation:

- **Simulation Acceleration [16]** – With this mode, synthesizable constructs are implemented onto the FPGA hardware, while non-synthesizable (Testbench) constructs remain in the RTL simulator. Simulation acceleration is a signal-based interface, as data is passed serially through the high speed channel. This mode is faster than RTL simulation, if the majority of the simulation time is not spent inside the testbench. If the testbench contains a majority of the simulation time, then the emulator will spend a most of the time in the communication structure between the hardware and software.
- **Transaction-based Emulation [17]** – In transaction-based emulation, the communication channel utilizes transactions, as opposed to signals in simulation acceleration. This is accomplished with high level testbenches on the software side and bus functional models (BFMs) which translate high level messages to low level signals which are the input to the DUT. The translators are commonly referred to as transactors, and typically implemented as finite state machines (FSMs), which send many signals to the DUT based on a single message. In this manner, a single message from the high level test bench can equate to hundreds of clock cycles in hardware, and conversely, hundreds of clock cycles in hardware can equate to a single message when sent to the software. The Accelera Systems Initiative, the creators of SystemC, developed the standard co-emulation modeling interface (SCEMI), which defines a model for simulations to run in an emulation environment and vice versa. Details of this SCE-MI interface are provided in the next Section.

### 2.4 Virtual Prototyping

Virtual platforms allow software engineers to model system-level behavior using TLM models. These high-level models can be paired with the development of a software stack, which can model the behavior of a system at real-time speed. Today's virtual platform systems have a large portfolio of operating systems, peripherals, processors, bus models, and various blocks which are commonly found in SoC devices. Software engineers can utilize the TLM models for customer demos which can present an early reference model prior to hardware development. The disadvantage of using a virtual prototype is that they are not a good representation of cycle accurate hardware behavior.

For scan design, the virtual prototypes utilize high level models, so a high-level model for a serial communication device (ex. JTAG) can be directly connected to the scan chain. Engineers can utilize C/C++ debugging tools, such as the GNU Debugger (GDB), to observe the software stack while signals transition in and out of the scan chain. Since these are high-level models, and will not be implemented into an FPGA, they are only representative of the hardware functionality, and not its implementation behavior.

## 3. STANDARD CO-EMULATION MODELING INTERFACE

SCE-MI implements a communication infrastructure which allows messages (transactions) to be passed between high-level software models to the device under test implemented in hardware. There are three types of SCE-MI interfaces: macro-based, function-based, and pipes-based. In this study, we will be using function-based and omit the other two interfaces. There are three main environments when describing the SCE-MI macro-based interface: the hardware side, the software side, and the SCE-MI bridge. On the hardware side, SCE-MI defines a set of synthesizable message ports which relay messages to and from the software side. The transactor is a bus functional model, which translates high level calls from the software side to bit sequences for the DUT. The SCE-MI Infrastructure (Fig. 2) also contains dedicated clock and reset control logic to be able to control system clocks.

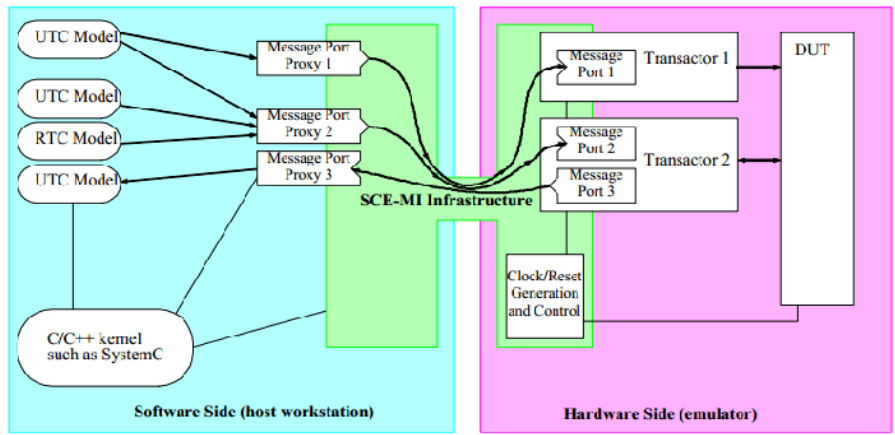

Fig. 2 SCE-MI Infrastructure.

The software side on the host workstation contains a set of message port proxies, which are implemented as C++ objects to allow the SCE-MI API to access the channel. From the connection to the channel, the proxy can connect to any untimed C model (UTC). The SCE-MI bridge utilizes a dual –ready handshake, in which software proxies and hardware message ports use Receive/Transmit ready signals to inform the other side that it is ready to receive or send data. The bridge channel acts as a bi-directional network socket, which carry the message, but it is the transcator's responsibility to deliver cycle-accurate information to the DUT.

## 4. METHODOLOGY AND TEST PLAN

To facilitate discussion of the proposed verification methodology, we will be focusing on ISCAS89 S400 sequential benchmark circuit only. It should be noted that the ideas presented here are generic that they are applicable to help verify any other large scale, scan-based designs.

### 4.1 Design Under Test – ISCAS S400 Benchmark

The S400 is a netlist description of a traffic light controller, which contains 21 registers and 3 primary inputs and 6 primary outputs [20]. The circuit is built with 58 inverters and 106 gates (11 ANDs + 36 NANDs + 25 ORs + 34 NORs). The 400 in the circuit description represents the number of interconnect lines among the circuit primitives. Amongst the primitives is a low-level description of a register module built using inverters, tri-states, and NMOS transistors.

After the scanned version of S400 is defined, all 21 scan registers are daisy-chained to another, with the first input and last output connected to test data in and test data out respectively. In addition, each register will have a reset signal that can reset the register to a known state prior to the scan sequence. A 'ScanEnable' signal is also added to each register to be able to put each register into test mode or allow normal functionality when de-asserted.

### 4.2 Test Bench & Plan

The testbench will verify the functionality of the circuit during RTL simulation. Prior to creating the testbench, a test plan needs to be created to exercise all modes of operation. Since we are primarily focusing on scan chain operation, we will need to create multiple test sequences for the DUT. The four test sequences which will be used are the reset toggle, scan enable toggle, scan sequence, and clock generation.

The first two test sequence puts the DUT into a known state, and enables operation of the scan chain. For the reset toggle, an active-high reset signal initializes all scan registers with a '0' value, allowing the DUT to clear any data which may be held during circuit initialization. During the testbench initialization, the reset is set low and is asserted/de-asserted prior to the scan enable toggle sequence.

The scan enable toggle sequence allows the testbench to shift in test vectors serially into the test data in port of the first scan register. The scan enable de-asserts when the vector is completely shifted into the scan-chain, capturing the output of the combination logical back into the scan chain registers. When the capture is finished, the scan enable is again asserted, which shifts out the current vector in the chain while simultaneously shifting in the next vector in the test sequence.

The scan sequence serially shifts in test vectors in the DUT when the scan enable toggle sequence is occurring. For the scan sequence, we will exercise full-scan mode, which utilizes all 221 test vectors. The counter is constrained to 221 vectors, and when it reaches the upper-bound, the testbench automatically exits following the last vector being shifted out. The clock generation generates the clock pulse, which feeds the scan registers and controls the speed of the DUT. For this simulation we will set the clock speed to 100MHz with a force command in the simulators debug properties. The main focus during RTL simulation is to determine where the testbench is spending a majority of the simulation time in the testbench. Although the clock speed can be adjusted, the CPU will throttle the performance based on current processes occurring, CPU multi-thread capabilities, CPU cores available, etc.

### 4.3 RTL Simulation & Profiling

The modified ISCAS S400 netlist will be first simulated with the Aldec Riviera-PRO functional verification platform [21]. Test bench and test sequence will reside in the workstation CPU. RTL simulation serves as the median to verify circuit operation, and allow debugging using the waveform viewer tool. The testbench based on the test sequence in the prior section (full-scan) is implemented on an HP Laptop with an Intel Core Duo CPU clocked at 2.13 GHz. To be able to benchmark the simulation run-time, Tcl scripts are created with processes which evaluate system time at the beginning and end of the simulation.

Two types of HDL simulation will be run: one run without profiler disabled and one with profiler enabled. The disabled profiler run allows us to verify functionality and ensure the circuit is operating correctly, but does not allow us to see the communication bottleneck between the DUT and the testbench. Since there are $2^n$ possible vectors to be fed as the primary inputs, profiler information is needed to assess the possible speedup that can be attained in simulation acceleration.

### 4.4 Simulation Acceleration

To reduce the number of events occurring in the CPU, the DUT will be transferred to the FPGA hardware, and will connect physically to the workstation via PCIe connector. The PCIe connector will facilitate information serially between the hardware and simulator via a co-simulation interface. With this methodology, there are two portions of the HDL code: synthesizable and non-synthesizable.

**Table 1** Synthesizable vs. Non-Synthesizable Logic

| Synthesizable (Implemented on FPGA) | Non-Synthesizable (Remains in HDL Simulator) |
|---|---|
| Ports | Delay statements |
| Signals and variables | Device initialization |
| Procedures | Assign statements |
| Modules | User defined primitives |
| Functions | Force and release |

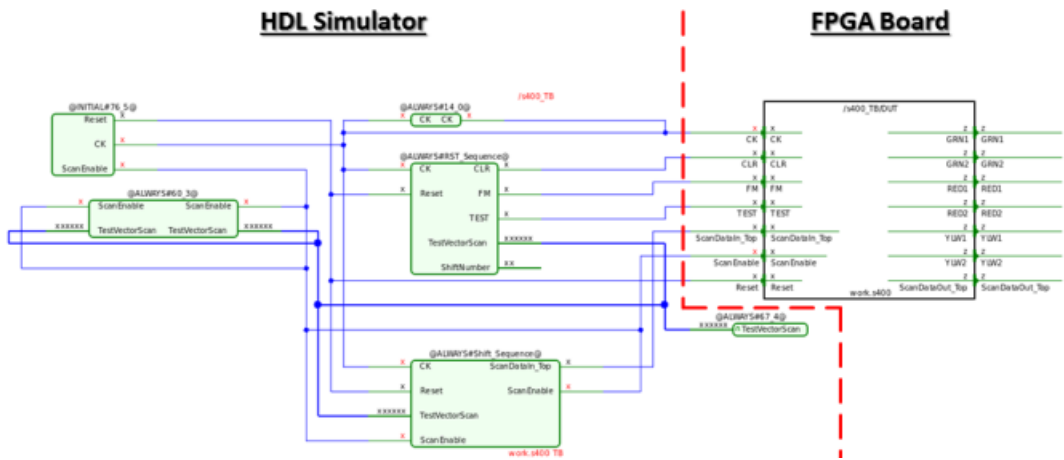

Fig.3 Splitting RTL Simulation Environment for Simulation Acceleration.

Figure 3 displays the separation between the system components which are non-synthesizable and synthesizable. The emulation compiler, Aldec Hardware Emulation Solutions Design Verification Manager [22] used in this study, automatically analyzes RTL sources and distinguishes between code which will remain in the simulator, and code which will be implemented onto HW. The DUT (S400) is fully synthesizable and implemented to FPGA LUTs, while the testbench constructs remain in the HDL simulator. For this design, the non-synthesizable constructs include Clock generation, Reset sequence, Scan Sequence, and Scan Enable Sequence. The communications between the HDL simulator and DUT are signal-based, meaning a single bit is transferred over a single-ended (SE) IO line available on the FPGA. The design has 7 inputs and 7 outputs, requiring a total of 14 SE IO, but the speed of the simulation is dictated by the data transferring on these lines.

The complete process for simulation acceleration is shown in Fig. 4 [23]. The design import stage of acceleration allows importing the FPGA design libraries or simulation libraries which are used during RTL simulation. Each file in the library in analyzed by the emulator, and automatically determines the synthesizable and non-synthesizable code. After the code is analyzed, the user selects a top-level instance, and configures the emulation options in the second stage. A user can instrument debug probes, partition design instances in multiple FPGAs, and synthesize the DUT with selected options with FPGA vendor tools. After running a place-and-route process, which physically maps the design to FPGA primitives, scripts are automatically generated which instantiate the communication between the hardware and simulator.

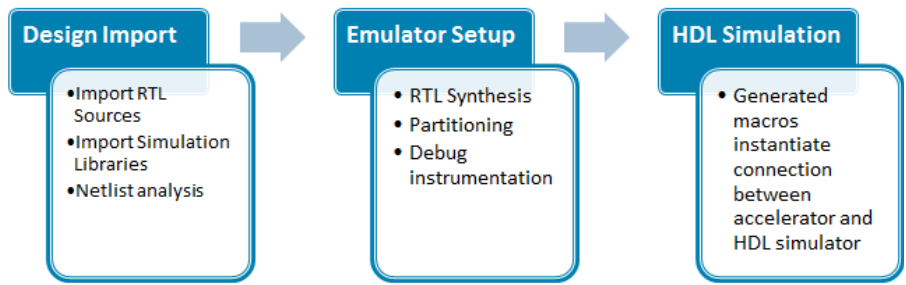

**Fig. 4** Emulation Setup Flow [23].

## 4.5 Transaction-Based Emulation

To utilize transaction-based emulation, modification needs to be done on the hardware and software side. The pre-existing testbench cannot be used, since the interface SCE-MI interface uses transactions for communication, as opposed to signals in simulation acceleration.

### 4.5.1 Software Side Modifications

The biggest change of the test environment will be converting the testbench sequences to high-level SystemC constructs. Whereas the RTL testbench relied on events such as the rising edge of a clock, the SystemC testbench focuses more towards the implementation of the sequence, rather than its trigger events. This will be done through two primary functions: timulus and read. The purpose of the stimulus function is to generate all the test sequences defined from the RTL testbench. Using C/C++ constructs, they are converted to high-level implementation. The sequences are converted as follows:

*Clock Generation* –Since the clock itself is required for hardware operation, two clock-related ports must be defined to operate correctly: clock port and clock control. These two ports are directly synthesized onto the FPGA, and have parameters which can be modified for multiple clocks, positive/negative edge triggers, duty cycle, and phase.

*Reset Toggle* –The reset signals are defined in the clock control and clock ports, which are directly synthesized onto the FPGA. User can modify HDL parameters to define the length of the reset sequence.

*Scan Enable Toggle* – The scan enable toggle sequence is accomplished by generating a 7-bit vector (similar implementation to RTL simulation/acceleration), and setting the scan enable input high. For the TDI value, a value of 0x1 is shifted in after the reset. The data in the scan chain is shifted out, and verified when received by the results function in the testbench. After the reset sequence, all the scan registers should have a value 0x0, and by shifting a value of 0x1 into chain, it verifies all registers can change states correctly. The stimulus function outputs are summarized in Table 2.

**Table 2** Test Sequences

| Test Sequence | Stimulus Function | Results Function |
|---|---|---|
| Generation | N/A<br>(SCE-MI defines system clocks) | N/A<br>(SCE-MI defines system clocks) |
| Reset Toggle | N/A<br>(SCE-MI defines Reset ) | N/A<br>(SCE-MI defines Reset ) |
| Scan Enable Toggle | -Scan Enable = 0    1<br>-Test Mode = 0    1<br>-Primary IO = X<br>-Send testbit '1'<br>-Scan Enable= 1    0 | Read output vector from message proxy, and display to output string construct. Verify bit vector is of value '1'. |
| Scan Sequence | -Scan Enable = 0    1<br>-Test Mode = 0    1<br>-Primary IO = X<br>-input all bit of first test vector serially to TDI pin with FOR loop<br>-Scan Enable = 1    0<br>-Increment test vector<br>-Loop last 3 steps for all $2^n$ vectors | Read output vector from message proxy, and display to output string construct. |

International Journal of Computer Science & Information Technology (IJCSIT) Vol 6, No 4, August 2014

*Scan Sequence* – After the scan enable sequence, the shift register has been verified to switch between the 0 state (reset sequence) and 1 state (scan enable sequence). For this SystemC testbench, there is a 7-bit message which will be sent to the hardware side. One of the inputs is a dedicated scan in port, which a 21-bit test vector will be serially shifted in. Figure 4 on the left below shows the shift and write sequence controlled by a FOR loop construct. The data is first shifted into hardware, and written to the message port. In the next section, an FSM is added to decrease the number of write tasks to the hardware.

### 4.5.2 Hardware Side Modifications

*Transactor core* – The transactor core receives the transactions from the software side (inMessage) and sends them to the DUT as signals. For this implementation, the SystemC testbench controls the shift sequence with a FOR loop, so data from the transactor core is passed and connected directly to the DUT. This is sometimes referred to as a 'dummy' transactor (Fig.5), since there is no manipulation of data in the transactor core. In the next section, we will describe an FSM implementation in HW, which will decrease communication for bit shifts into the DUT.

```
elsif(rising_edge(Uclock)) then
    if(dataReady = '1') then
        out_xtor_to_DUT_GND_Reg              <= inMsgVect(7);
        out_xtor_to_DUT_VDD_Reg              <= inMsgVect(6);
        out_xtor_to_DUT_FM_Reg               <= inMsgVect(5);
        out_xtor_to_DUT_TEST_Reg             <= inMsgVect(4);
        out_xtor_to_DUT_CLR_Reg              <= inMsgVect(3);
        out_xtor_to_DUT_ScanDataIn_Top_Reg   <= inMsgVect(2);
        out_xtor_to_DUT_ScanEnable_Reg       <= inMsgVect(1);
        out_xtor_to_DUT_ScanTestMode_Reg     <= inMsgVect(0);
    end if;
end if;
```

Fig. 5 Transactor Pass Through Assignments.

*SCE-MI Message Ports* – The message port consists of an inPort and an outPort. The outPort sends data from the transactor to the software side, and the inPort receives data from the software side to the transactor core. The message ports utilize a dual ready handshake protocol with three primary I/O's: ReceiveReady, TransmitReady, and message.

When both the ReceiveReady and TransmitReady are asserted high, the message is sent to the destination across the channel. This occurs on the active edge of the Uclock, which allows the ports to be written and read to while DUT is still in operation. On the software side, when a message port is instantiated (whether it is in or out) in the testbench, a first-in-first-out (FIFO) memory is created where instructions are stored prior to entering and leaving the proxy. The instruction from the testbench is sent to the FIFO, where the proxy will wait for the ReceiveReady on the hardware side. The proxy will assert a TransmitReady (for an inPort) or ReceiveReady (for an outPort) when the FIFO contains data, and flush the data out, when ready on the hardware side. Fig. 6 is a sample of the C++ implementation of the SCE-MI InPort which is used in the SystemC TB.

International Journal of Computer Science & Information Technology (IJCSIT) Vol 6, No 4, August 2014

```
//Read input FIFO.
inData = InPort.read();

//Create data structure for port.
for(unsigned int wrd = 0; wrd < IN_PORT_SIZE/32; wrd++)
{
    scemi_msg.Set(wrd, (SceMiU32) inData.range( (32*(wrd+1)) - 1, 32*wrd).to_uint(), NULL);
}

// Wait for data port to be ready
InPortReady.wait();
PRINT_DEBUG << "Scemi Port: " << name() << " sends data: " << inData.to_string(SC_HEX) << std::endl;

//Send data.
InPortProxy -> Send(scemi_msg);
```

**Fig. 6** SCE-MI Message In-Port Software Implementation

***SCE-MI Clock Port*** – The SCE-MI clock port provides the DUT with a controlled reset and clock. Through a set of parameters such as clock duty, clock ratio, phase, and reset cycles, the user can customize how timing and reset is handled in the circuit. For circuits with multiple clock frequencies, multiple clock ports have to be instantiated with parameters customized for each clock frequency. Since the system used in this paper contains only a single global clock, one clock port and clock control is needed. If no parameters for the clocking are set, SCE-MI automatically generates a 1/1 ratio clock, a single clock with the highest frequency in the system. Most of the time, SCE-MI will grab this clock from oscillators available on the emulation board. Each EDA vendor provides the input clock for SCE-MI based on tests and delay analysis on a specific FPGA board. The end-user needs not to focus on the clock implementation in the HW, but only to create the necessary clock port, which will call the correct frequency built on the emulation API.

***SCE-MI Clock Control*** – The SCE-MI clock control macro is a macro which aligns clock edges from the uncontrolled clock and controlled clock. This clock control enables freezing the DUT controlled clock, while still operating the transactor to receive incoming data from the software side. The main advantage of this control module, which will be used in the modified core implementation, is the ability to perform operations on data while the DUT is still "frozen". ReadyForCclock is a signal in the control module that allows the DUT clock to advance. If data received from the software of the DUT needs to be analyzed, ReadyForCclock can be set low, which allows operation on the data from the DUT or software side to occur without the DUT clock running.

### 4.5.3 Overall SCE-MI Test Environment

Here the test vector will be generated in the software, but the entire vector will be within the message sent to the transactor core. This cuts out the serial processing of each bit to the inPort, and the number of writes/reads between the HW/SW interfaces. The message vector and signals will pass through the transactor core, which contains an FSM (Fig. 7) that will serially shift data in to the DUT.

84

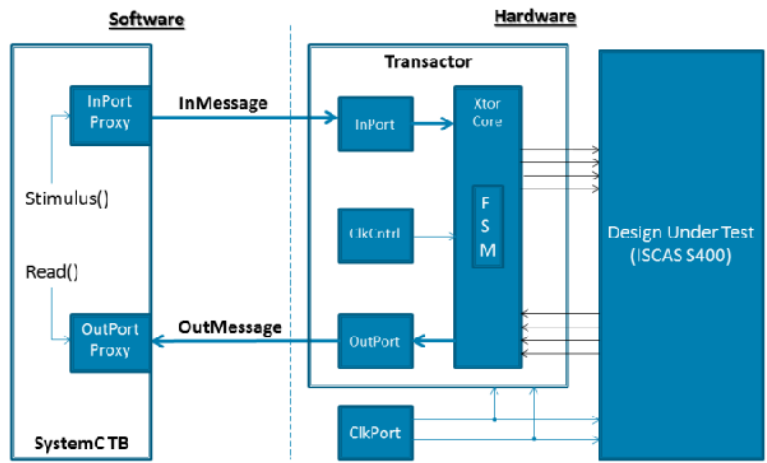

Fig. 7 SCE-MI FSM Test Environment.

The FSM machine (Fig. 8), which is regulated by Uclock, will contain multiple states for all inPort/outPort calls, vector reads, and shifting. As the FSM cycles through all the states, the scan control signals also regulate the flow of data to the scan chain. The inPort/outPort calls regulate the dual ready handshake protocol, which asserts appropriate signals before and after the shift sequence. After the shift sequence has finished, the output port transmits data, while new data is called to the inPort simultaneously. After new data is received, the shift sequence begins again. The FSM must be reset appropriately, so all the signals are initialized. The Ureset initializes all signals prior to the data exchange sequence between the hardware and software. This ensures that the FSM does not begin in an unknown state.

```
FSM:process(Ureset, Uclock)
begin
    if(Ureset = '1') then
        out_xtor_to_DUT_FM_Reg              <= '0';
        out_xtor_to_DUT_TEST_Reg            <= '0';
        out_xtor_to_DUT_CLR_Reg             <= '0';
        out_xtor_to_DUT_ScanDataIn_Top_Reg  <= '0';
        out_xtor_to_DUT_ScanEnable_Reg      <= '0';
    elsif(rising_edge(Uclock)) then
        if (CS = InPortCall) then
            CS <=   VectorReceived;
            OutportReady2Send  <= '0';
            InportReady2Receive <= '1';
        elsif (CS = VectorReceived) then
            CS                                  <= Shift0;
            InportReady2Receive                 <= '0';
            OutportReady2Send                   <= '1';
            out_xtor_to_DUT_FM_Reg              <= inMsgVect(23);
            out_xtor_to_DUT_TEST_Reg            <= inMsgVect(22);
            out_xtor_to_DUT_CLR_Reg             <= inMsgVect(21);
            out_xtor_to_DUT_ScanEnable_Reg      <= '1';
        elsif (CS = Shift0) then
            CS <= Shift1;
            out_xtor_to_DUT_ScanDataIn_Top_Reg  <= inMsgVect(0);
        elsif (CS = Shift1) then
            CS <= Shift2;
            out_xtor_to_DUT_ScanDataIn_Top_Reg  <= inMsgVect(1);
```

Fig. 8 Transactor's FSM Transitions.

## 5. RESULTS & ANALYSIS

The FPGA-based prototype solution which is adopted in this study is the Aldec HES5XLX660EX, which features two Xilinx XC5VLX330T FPGAs for a total capacity of 5 million ASIC gates. Users are able to connect the HES5 board to a host workstation via PCI Express x8 lane communication. A dedicated FPGA provides host interface logic when connecting the Aldec HES-DVM emulator with the FPGA prototyping board.

### 5.1 RTL Simulation Results

A Verilog testbench is generated for the S400 DUT, with processes executing each of the test sequences. The testbench is roughly ~150 lines of code, and includes the instantiation of the DUT module. Inside the testbench is a 21-bit count register, which increments after a test sequence is fully shifted into the scan chain. After the initial reset sequence, the count value is serially shifted into the test data input from least significant bit (LSB) to most significant bit. When the test vector is shifted in, a flag to stop the shift is asserted, and the capture process begins. The data in the scan chain passes through the gate-level logic, and is captured back into the scan registers. The data is then shifted out with the same Verilog process that shifts in the data. For all the Verilog files, a timescale of 10ns period is used (i.e. a 100MHz clock frequency).

Using Aldec Riviera-PRO functional verification tool, compilation and simulation macros/scripts are created to automate the simulation process. To calculate the runtime of the simulation, a TCL script can capture the time of the CPU before and after the simulation run, and determines the difference between the two times. The time difference is then saved to text file, which can be observed after the simulation has finished. Riviera-PRO has a TCL-based console, which allows the timer script to be integrated into the macro which runs the compilation and simulation. The timer process begins once the design is compiled and elaborated. The elaboration time is the time required by the simulator to process the RTL source files, and establish the events queue. This time will not be included in the final run-time calculation, since it is EDA vendor dependent, and can differ depending on how the simulator processes the RTL code. After running through Riviera-PRO, the benchmark time for RTL simulation was 921 seconds. The sequence ends with the Verilog $finish command after the last test vector is scanned out and saved to the text file. The test vectors will be later used as the baseline (a.k.a, golden sequence), when the test environment is ported to simulation acceleration and emulation.

The next step is a simulation run with profiler to determine which portions of the test environment (DUT + TB) take a majority of the total simulation time. To activate profiler, Riviera-PRO uses special debug switches during compilation, which analyzes the design structure for debug purposes. This analysis, however, requires additional effort by the CPU to analyze the data since there is a large amount of data compared to a simulation run without profiler. Typically, design profilers or debug capabilities are meant for regression testing since they elongate the simulation run-time. After including the additional debug switches to the compilation scripts, the design was re-run with the data shown in Table 3. The original simulation time of 921 seconds increased to 1346 seconds when profiler was activated, a difference of 421 seconds compared to the original run.

Table 3  RTL Simulation Results

|  | Simulation Time (s) |
|---|---|
| (DUT + TB) | 921 |
| RTL Simulation w/ Profiler  (DUT +TB) | 1346 |
| (DUT + TB) | 921 |
| **Total Profiler Time** | 421 |

The profiler calculates CPU ticks for each process accessed during the test time, along with all sub-processes which are called. Clock generation is a continuous process which runs in sequence of all events of the design. Since the DUT operates on clock generated sequences, the clock process took a majority of the CPU resources during the simulation. Withholding the clock generation sequence gives us a better understanding of the test sequences that directly correlate to the scan chain sequence. The profiler reported the following results for simulation time percentages (Table 4). The TB and DUT had near equal amounts at 44.71% and 55.29%, respectively. Of the testbench processes, the reset sequence totaled 1.2%, the scan enable sequence 7.3%, and 36.21% for the scan sequence. The largest portion of the testbench is circulated around the scan sequence, since it requires running all the test vectors through the scan chain serially. The DUT alone takes about 55.29% of the simulation time, which correlates to the shifting between scan registers and the combinatorial datapath. With the profiler results, we can estimate the amount of speedup which we can achieve with simulation acceleration.

**Table 4.** RTL Profiler Results

| Module | Simulation Time (%) | Simulation Time (s) |
|---|---|---|
| **TB** | 44.71 | 411.78 |
| Reset Sequence | 1.2 | 11.05 |
| Scan Enable Sequence | 7.3 | 67.23 |
| Scan Sequence | 36.21 | 333.49 |
| **S400 DUT** | 55.29 | 509.22 |

To estimate the speedup for acceleration, we need to use the percentage of the HDL simulator time spent in the testbench. This portion of test environment includes all non-synthesizable HDL logic which cannot be implemented on FPGA fabric, and must remain in the HDL simulator. Since the testbench is driver logic for the DUT, it will need to communicate the signals over the physical link to the acceleration platform. If the majority of the acceleration time is spent in the testbench and physical link, the speedup factor will not be large. The relationships between the two are indirectly proportional as the simulator time % increases the speedup factor decreases. With a testbench percentage at 44.71%, the simulation acceleration factor can be estimated at about 2.23x faster than RTL simulation. This would mean with the testbench implemented in the HDL simulator and the DUT synthesized onto FPGA logic, we can expect to see a simulation time of roughly 440-450 seconds with simulation acceleration.

### 5.2 Simulation Acceleration Results

With simulation acceleration, the existing test environment can be reused without any modification. Prior to running simulation acceleration, we need to setup the test environment with Aldec HES emulation software. The emulator process takes in RTL sources files and splits all synthesizable and non-synthesizable logic. The synthesizable logic is then translated into FPGA flip flop and LUT primitives who are mapped onto FPGA fabric. The DUT itself can be manually separated and partitioned amongst separate FPGAs, but since we are targeting speed as the end result, it is best to place all logic into a single FPGA. The emulator then scans the design hierarchy, and provides a report on all the resources used when the design is translated using Xilinx synthesis tool.

The DUT portion requires only 21 of 207360 flip flops (scan registers), and 53 of 207360 look-up tables of the Xilinx Virtex-5 FPGA. Since the DUT will be synthesized onto hardware, the signals within the design will not be observable. To be able to debug the design, additional instrumentation needs to be implemented for each signal the user chooses to debug. The number of signals needed to debug is directly proportional to the amount of logic added by the emulator.

Since we need to observe all primary IO and internal signals which connect all scan chains, we mark all the signals and IO as 'static debug' signals. During the acceleration run, the debug data is captured with additional registers, and can be analyzed within the RTL simulator. In addition to debug instrumentation, additional logic is required for the interface logic and the emulation controller. The controller interface provides the necessary signals to be able to connect to the host workstation over the physical channel. Aldec HES-5 board comes with an additional 'interface FPGA' which houses this emulation interface, so user logic availability is not reduced [16]. After user chooses debug and partition options, the design is resynthesized with the logic added into the netlist. Prior to place and route, the emulator provides an updated resource report which includes the DUT and emulator inserted logic. In Table 5, we see that the emulator inserted 81 additional registers and 10 additional look-up tables for the debug logic. This equates to a flip flop percent increase of 385.71% and 18.86% for look-up tables. We find that, although simulation acceleration provides an increase in speed, it requires additional area to be able to debug signals.

**Table 5.** Synthesizable Logic with Debug Resources

| Synthesizable Logic (DUT + Emulator Logic) | | | |
|---|---|---|---|
| **Resources** | **Amount Used / Amount Available** | **Additional Resources Required by Emulator** | **% Increase** |
| Flip Flops | 102/ 207360 (0%) | 81 | 385.71 |
| Look-Up Tables | 63/207360 (0%) | 10 | 18.86 |

After the place and route process, a bit file is generated, along with the necessary scripts to instantiate the acceleration. The scripts just need to be integrated into the design directory, which in this case, is the original simulation directory. After connecting the scripts to the design directory, the simulation acceleration is run. After the run, the benchmark text file is observed for the new simulation time, in addition to ensuring that the golden vectors are matched correctly from the original simulation run. Compared to the original RTL simulation time of 921 seconds, the simulation acceleration time is 435 seconds, a time difference of 486 seconds faster than RTL simulation.

Looking at the 2.23x expected speedup determined by the profiler, we attained a speedup value of 2.11x during the acceleration run. The delta between the two speedup values is due to the PCIe physical channel, in which signals have to traverse as they pass from the hardware side to the software side. If the number of signals is large, and there is constant communication between the two interfaces, the speedup factor can decrease. Using the profiler tool with the simulation acceleration run, we can see that the PCIe physical link takes up 10% of the simulation run, as the other 90% is testbench logic. Since the DUT portion is no longer computed by the CPU, there are less instructions and events to process by the simulator, thus speeding up the simulation time.

### 5.3 Transaction-Based Emulation Results

#### 5.3.1 Pass-Through Transactor

To accomplish transaction-based emulation, changes needed to be made to the testbench and RTL. A SystemC testbench is created which generates a test vector, and serially shifts it to the DUT via message ports implemented onto the FPGA. In addition to the message ports, clock ports and clock control modules are implemented onto hardware which control the system clock and reset. The top-level portions of the design which will be implemented with the emulator are the DUT, transactor core, and clock ports. The transactor core instantiates the message ports and clock control modules, as well as the logic which translated the top-level message to low-level signals which will be sent to the DUT. Prior to implementation on hardware, Aldec provides a SCE-MI simulation environment, which allows users simulate SCE-MI modules with the DUT prior to moving to hardware. In this test environment, the testbench is a SystemC executable, while the hardware portion is simulated in the RTL simulation. After verifying that transactions are correct in the waveform, we can move the design into the emulation tool.

Design import in the emulator tool automatically recognizes SCE-MI module definitions, and sets up the proper environment variables for emulation. Since the emulation platform will contain the SCE-MI modules, transactor logic, and the DUT, the area is much greater as compared to simulation acceleration. The area used for emulation with the pass through transactor is 64 flip flops and 97 LUTs prior to implementing any debug logic. This is 43 flip flops and 44 LUTs greater than simulation acceleration.

With emulation, there are two options for debugging. Dynamic debugging is an option which utilizes Xilinx readback, which reads FPGA registers dynamically during run-time, allowing the user to pause the emulation run [24]. The advantage is that debug instrumentation is lessened, since debugging it done through a Xilinx standard process. The disadvantage is that the readback process can be long, if the placement tool spreads the logic throughout the FPGA fabric. Another option is to utilize the debugging process used in simulation acceleration, which instruments more logic, but doesn't affect the simulation time as much as the readback. For this process, we use the later process, and instrument additional logic, since the design is fairly small. With additional debug probes, the logic grows to 102 flip flops and 143 LUTs, a percent increase of 59.37% and 47.42%, respectively, compared to no debug logic added.

The open source SystemC initiative (OSCI) and TLM standards can be downloaded from the Accelera Systems Initiative website [11]. This file contains all the required headers and C++ files required to process SystemC files. After being downloaded, a SystemC testbench can be processed using any Linux terminal in a similar fashion processing a C/C++ file. The SystemC testbench cycles through all the tests described in earlier sections, and use the terminal console to display messages to the user. To determine the simulation time, the testbench captures the time before and after the simulation, in a similar fashion to simulation acceleration. The HES emulator also provides two frequencies: the system frequency and the DUT frequency. The system frequency is the uncontrolled clock which feeds the transactor. Uncontrolled clock runs freely, and if the user were to integrate a software application, driver, or operating system, this is the frequency it would be able to operate. The DUT frequency is the controlled clock, which feeds directly to the registers in the scan design. It is worth to note that since the uncontrolled clock is running freely, it is generally faster than the DUT frequency. For this measure, we will look at the simulation run-time, DUT frequency, and number of reads from the software side from the hardware side. The number of reads from the testbench is the data being passed directly from the SystemC testbench to the hardware. The emulator does some internal packing of vectors being passed from the software to hardware, so there may be optimizations done on the read cycles occurring.

The SystemC testbench is started from a Linux console with the scripts generated by the emulator and the XML file. For the pass through transactor implementation, the total emulation time is 379 seconds, not including the time to download the bit file to the FPGA. The system frequency (uncontrolled) clock is 8.33 MHz, and the DUT frequency is 254.75 KHz with a read value of 3,241,936. The transactions being read back by the SystemC testbench analyzes each vector, and saves them to a text file in a similar fashion the golden vectors are used in RTL simulation. Prior to benchmarking the emulation, the vectors were verified against the golden vector for correctness.

Comparing the emulation run time to RTL simulation and simulation acceleration, emulation is 2.43x faster than simulation, but only 1.14x faster than simulation acceleration with a difference of 56 seconds. This is due to the structure of the transactor core, since the pass-through transactor serially shifts in data from the testbench to the DUT, a similar test environment as simulation acceleration. Compared to RTL simulation with a time difference of 542 seconds (Table 6), verification of the scan chain has been reduced by 58.8%. For large designs, this percentage can equate to hours or even days based on the complexity of the DUT. To optimize the emulation results, the transactor core needs to be modified in a way that reduces the number of reads from the SystemC testbench. After some vector compacting and optimizations to the proxy and message ports by the emulator, there is a total of 3,241,936 reads from the hardware side to the software side. Since the hardware side has to read the data from the software during each clock cycle, the number of reads can be large. This value is the bottleneck between the hardware and software in a similar manner the testbench is in simulation acceleration. Reducing this number will increase the bandwidth between the two interfaces, and allow for a faster emulation.

Table 6. Pass Through Transactor Emulation Comparison

|  | Simulation Time (s) |
| --- | --- |
| RTL Simulation | 921 |
| Simulation Acceleration | 435 |
| Transaction- Based Emulation (Pass Through Transactor) | 379 |
| **Time Difference between Emulation and Simulation Acceleration** | 56 |
| **Time Difference between Emulation and RTL Simulation** | 542 |

### 5.3.2 FSM Transactor

The FSM transactor core is a modified transactor which incorporates an FSM to serially shift in data sent from the software side. The software side sends a complete 21-bit test vector, and the transactor core has states which send each bit to the DUT. All control signals such as the scan enable and message calls to the software are maintained by the transactor FSM. The modifications are done on both the software and hardware side prior to porting the source files into the emulation setup tool. Compared to the pass-through transactor in the previous approach, the FPGA area is larger since it incorporates more logic. For the amount of synthesizable logic, the FSM transactor implementation uses 148 flip flops and 206 LUTs. Compared to the pass-through transactor implementation, the FSM transactor uses 84 additional flip flops and 109 additional LUTs. For debug instrumentation, the same signals probed for debug in the pass-through are used in the FSM implementation. This allows reuse of scripts which initialize all debug instrumentation. With the debug logic added, the total area used is 193 flip flops and 264 LUTs, an increase of 30.40% for flip flops and 28.15 for LUTs.

Re-running the emulation with the FSM transactor yields the following results (shown in Tables 7 and 8). The total emulation time is 158 seconds with a system frequency of 8.33 MHz. The DUT frequency is 681.84 KHz, and has a read value of 1383556. Comparing both emulation implementations, the FSM transactor is 221 seconds faster (2.4x speedup). Since the read value has been reduced by a value of 1858380 reads (57.23% decrease), the transactor does not have to freeze the controlled clock to the DUT as much as the pass-through transactor implementation. This allows the controlled clock feeding the DUT to run a faster clock rate. The FSM transactor implementation runs at 681.84 KHz, while the pass-through transactor runs at 254.75 KHz, a

90

International Journal of Computer Science & Information Technology (IJCSIT) Vol 6, No 4, August 2014

difference of 427.09 KHz (167.65% increase).The system frequency is dictated by on-board oscillators, which are set by emulation vendors based on board IO skew, internal testing, and worst case scenarios. Since the system frequency remains the same at 8.33 MHz for both emulation implementations, the oscillators must be locked at 8.33 MHz as a ceiling value for each FPGA on the HES-5 prototyping board.

Table 7. SCE-MI FSM Transactor Results

| Description | Value |
| --- | --- |
| Emulation Time | 158 seconds |
| System Frequency | 8.33 MHz |
| DUT Frequency | 681.84 KHz |
| Read Value | 1383556 |

Table 8. SCE-MI Emulation Comparisons

|  | Pass-through Transactor | FSM Transactor | Difference |
| --- | --- | --- | --- |
| **Emulation Time** | 379 seconds | 158 seconds | 221 seconds |
| **System Frequency** | 8.33 MHz | 8.33 MHz | - |
| **DUT Frequency** | 254.75 KHz | 681.84 KHz | 427.09 KHz |
| **Read Value** | 3241936 | 1383556 | 1858380 |

### 5.4 Complexity Analysis

The complexity of scaling scan chain designs is dictated by a complexity of . Scan methodologies such as partial-scan works to reduce the complexity by creating a subset of the test vector set to implement the scan chain versus all possible combinations. Since we are using a full-scan methodology, we must work on reducing the shift-in/shift-out process since we will be using all possible test vectors. In this study, we optimize the shift-in/out process by using various RTL simulation and hardware verification, but each of these methodologies has a specific bottleneck.

RTL profiler showcases how the CPU handles each process in the test environment which includes the testbench and the DUT. The testbench took up 44.71% of the total complexity, which a majority of the time was in the scan sequence. If we apply the complexity algorithm for the S400 scan-chain DUT, we incur a complexity of $O = [2^{21}(21+1)] + 21 = 46137365$ clock cycles. Applying the profiler results to the complexity, the testbench takes about 20,628,015 clock cycles, which will remain in the RTL simulator environment. The other 25,509,349 clock cycles will be implemented onto FPGA primitives in the hardware emulator. As the testbench percentage increases, more clock cycles will remain in the testbench, increasing the time of the simulation. Table 9 shows that as the testbench percentage increases, the larger the number of clock cycles implemented in the HDL simulator.

Table 9. RTL Simulation Clock Cycle Workload

| Testbench Percentage | Clock Cycles Implemented in TB |
| --- | --- |
| 80 | 46137365*.8 = 36909892 |
| 60 | 46137365*.6 = 27682419 |
| 40 | 46137365*.4 = 18454946 |
| 20 | 46137365*.2 = 9227473 |

SCE-MI provides a run-time API which allows users to probe the controlled and uncontrolled clock. The controlled clock is the number of running cycles in the DUT, while the uncontrolled clock is the free-running clock dictated by the board hardware. In the SystemC testbench, a debug function was created to probe the number of controlled clock sequences for the scan sequence. The results for the controlled clock reading for both SCE-MI runs are reported in Table 10.

Table 10. SCE-MI Controlled Clock Cycle Results

| SCE-MI Run | Controlled Clocks Required for Test |
|---|---|
| Pass-Through | 25211651 |
| FSM | 8238807 |

FSM implementation controlled the scan sequence within the transactor, so the serial shifting which took a majority of the test time, was handled outside of the DUT. We can see that comparing the FSM to the overall complexity of the scan chain, that the FSM transactor runs 5.6x faster than the estimated 46,137,365 clock cycles. The pass through implementation incurs a speedup of 1.71x, a smaller value due to the serial shifting from the SystemC testbench which connects directly to the DUT inputs from the registered data from the transactor. From this information, we can conclude that the RTL simulation and acceleration speedup is based on the complexity of the testbench, while for SCE-MI emulation, the speedup is dependent on the complexity of the transactor.

**5.5 Resource Analysis**

Transitioning from simulation acceleration to emulation, we see a steady linear growth of FPGA resources used. Although the S400 if a fairly small design, today's SoC designs push the max of FPGA resource utilization and are bound by a finite set of resources. Adding additional resources to the DUT may require a larger emulation platform with multiple FPGAs, which incurs a larger cost. Emulation capabilities such as debug, automatic partitioning, and memory interfaces generally are added by the EDA vendor through custom IP with a small (around 5-10% FPGA utilization) footprint. However, some emulation tools (ex. Debug probes) can have a large effect on FPGA resources, requiring the user to utilize other debug methodologies such as debug daughter boards and logic analyzers.

Fig. 9(a) shows the resource utilization (LUTs and Flip Flops) for each acceleration and emulation run. In simulation acceleration, only 21 FFs were required (scan-FFs) since we were reusing the existing test environment in RTL simulation. As we moved toward FSM emulation, we increased flip flop to 148, a growth of 127 flip flops. Since we are adding more resources onto the FPGA (transactor + SCE-MI modules) for emulation, the utilization grows depending on the complexity of the transactor. This is evident when comparing the pass-through emulation run with the FSM emulation run. Since the pass-through transactor just connected signals between the SystemC TB and the DUT, the amount of logic required was not that large. Compared to simulation acceleration, the pass-through transactor only required 64 flip flops and 97 LUTs, a difference of 43 flip flops and 44 LUTs. The main bottleneck when additional resources are utilized in emulation is the complexity of the transactor. Also, if the DUT requires multiple transactors to test separate functionality (ex. Transactor for ingoing and outgoing traffic to a USB port), then the number can increase.

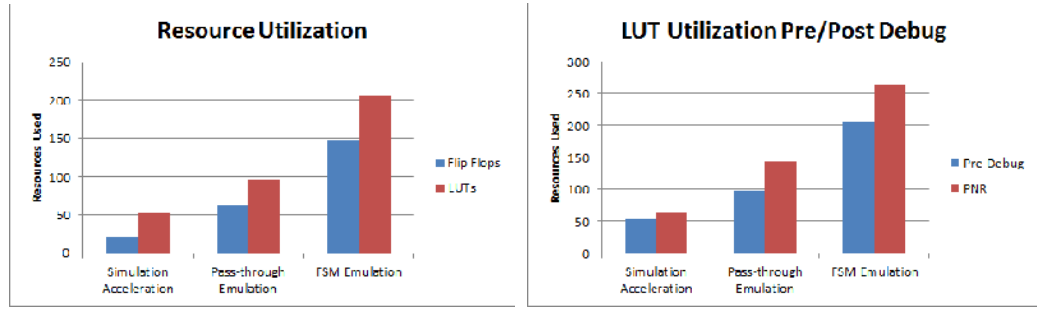

**Fig. 9(a)** Resource Utilization for Acceleration and Emulation (left); (b) LUT Utilization Pre/Post Debug Implementation (right).

Debug probes have a large effect on speed and FPGA resources utilized in simulation acceleration and emulation. To be able to bring design top-level ports or internal signals synthesized on FPGA into a waveform environment for debug requires additional logic implemented by the emulator. Since the emulator encases the DUT with a wrapper connecting the internal logic to the external interface – additional registers/LUTs need to be implemented to bring those signals to the top-level of the wrapper. If a large amount of signals need to be debugged, than more registers are requires. As the number of registers/LUTs increase, the number of events occurring between the hardware and software interfaces increases, having an effect on simulation speed.

Fig. 9(b) shows how the LUTs increase pre-debug and the final reported number in PNR. Prior to debug in the pass-through emulation run, the LUT count was 97, and grew to 143 (47.42% percent change) after debug signals were implemented. This increase was due to a large number of wide (32-bit) signals which were messages from both the inPort and outPort SCE-MI modules. In addition to DUT top-level signals, SCE-MI control signals (clocks, resets, etc) were also required to validate the incoming and outgoing transmission of data. Also, as the transactor complexity increased, more signals were required for debug to verify the transactor logic was correct after PNR occurs and module is implemented onto the FPGA. The FSM emulation run required 206 LUTs pre-debug, and 264 LUTs were reported during the PNR process. Since the transactor core was implemented with a state machine with multiple states controlling the scan sequence, the LUT count increase was greater than the other two runs. For the pass-through run, the LUT count increases by 46 while the FSM transactor increased the LUT count by 58 LUTs. As the complexity of the transactor logic increases, the number of signals required for debugging increases since the logic needs to be verified in conjunction with the DUT, that proper data is being sent and received. A strategy many verification teams utilize is inserting debugging probes only during regression runs to save on emulation time. The initial emulation is run, and if the data is found to be incorrect, then additional time and resources is utilized to debug the circuit. This saves time and FPGA utilization, since multiple designs can be verified in parallel implementing debug probes only when necessary.

## 6. CONCLUSION

In this paper, we showed how to utilize a SCE-MI infrastructure for testing scan-chain implemented designs on an FPGA-based prototype for reduced verification time and increased system performance. During RTL simulation, we verified the scan-chain ISCAS S400 benchmark circuit in 921 seconds, and while reusing the same test environment accomplished an acceleration time of 435 seconds. This 2.11x speedup from RTL simulation allowed us to maintain a level of visibility of internal signals utilizing the static debugging capabilities of the emulator. Moving to the transaction-based emulation environment, we saw an emulation time of 379 seconds for the pass-through transactor, and 158 seconds for the FSM transactor. Compared to the RTL simulation implementation, the FSM transactor was able to accomplish a 5.8x speedup, with a maximum system frequency of 8.33 MHz. With emulation, the FSM transactor was able to completely offload the scan-in and scan-out process to the hardware, decreasing the scan complexity by a factor of $2n$. The results of this study allow large scale scan-chain based SoC designs to be verified in a high-speed environment. Running the system in the sub-megahertz range, as accomplished in transaction-based emulation, also allows verification teams to integrate software early in the verification cycle.

## ACKNOWLEDGEMENTS

This material is based, in part, upon work supported by the National Science Foundation under grant numbers IIA-1301726 and CNS-1126688, and a University of Nevada Las Vegas (UNLV) Faculty Opportunity Award.

## AUTHORS

Bill Jason Tomas received his M.S. in electrical engineering from the University of Nevada, Las Vegas, in 2013. He is currently working at Cadence Design Systems, Inc.

Yingtao Jiang received his Ph.D. in Computer Science from University of Texas at Dallas in 2001. He is currently a Professor in the Department of Electrical and Computer Engineering at the University of Nevada Las Vegas. His research interests include VLSI, algorithms, computer architectures, biomedical signal processing and instrumentation, wireless communications and security, and nanotechnology.

Mei Yang received her Ph. D. degree in computer science from the University of Texas at Dallas, TX, in 2003. She is currently an Associate Professor in the Department of Electrical and Computer Engineering at University of Nevada, Las Vegas, NV. Her research interests include computer architectures, on-chip interconnection networks, embedded systems, and networking.